\documentclass{natureprintstylev4}

\usepackage{xcolor}
\usepackage{graphicx}	
\usepackage{amsmath}	
\usepackage{amsmath}	
\usepackage{bm}
\usepackage{hyperref}
\usepackage[capitalize]{cleveref}
\usepackage[normalem]{ulem} 
\usepackage{array, float, adjustbox} 
\usepackage{multicol}

\usepackage{pdflscape}
\usepackage{afterpage}
\usepackage{capt-of}
\newcommand{\aj}{Astron. J.}   
\newcommand{\apj}{Astrophys. J.}   
\newcommand{\apjl}{Astrophys. J. Lett.}   
\newcommand{\apjs}{Astrophys. J. Suppl. Ser.}   
\newcommand{\ao}{Appl. Opt.}   
\newcommand{\aap}{Astron. Astrophys.}   
\newcommand{\jcap}{J. Cosmol. Astropart. Phys.}   
\newcommand{\mnras}{Mon. Not. R. Astron. Soc.}   
\newcommand{\nat}{Nature} 
\newcommand{\physrep}{Phys. Rep.}   
\newcommand{\pasj}{Publ. Astron. Soc. Jpn}   

\newcommand{\ie}{\textit{i.e.}}
\newcommand{\eg}{\textit{e.g.}}

\newcommand{\kms}{\,{\rm km/s}}
\newcommand{\Mpc}{\,{\rm Mpc}}
\newcommand{\Mpch}{\Mpc/h}

\newcommand{\LCDM}{$\Lambda$CDM }

\def\Mpch{h^{-1} \mathrm{Mpc}}

\newcommand{\vnabdot}{\bm{\nabla} \cdot}
\newcommand{\snl}{\sigma_{\text{\tiny NL}}}

\newcommand{\vv}{\bm{v}}

\renewcommand{\indent}{\hspace*{24pt}}

\newcommand{\cmap}[1]{#1}

\newcolumntype{R}[2]{%
    >{\adjustbox{angle=#1,lap=\width-(#2)}\bgroup}%
    l%
    <{\egroup}
}
\newcommand*\rot{\multicolumn{1}{R{45}{1em}}}
\newcommand\redsout{\bgroup\markoverwith{\textcolor{red}{\rule[0.5ex]{10pt}{2.8pt}}}\ULon}

\title{Identification of Basins of Attraction in the Local Universe}

\author{
    A. Valade,$^{1,2,3}$\thanks{E-mail: avalade@aip.de}~
    N. I. Libeskind,$^{1}$
    D. Pomarède,$^{4}$
    R. B. Tully,$^{5}$
    Y. Hoffman,$^{6}$
    S. Pfeifer,$^{1}$
    and E. Kourkchi$^{5}$}

\date{Accepted XXX. Received YYY; in original form ZZZ}

\begin{document}
\onecolumn
\maketitle

\begin{affiliations}
$^{1}$ Leibniz Institut f\"ur Astrophysik Potsdam (AIP), An der Sternwarte 16, D-144 Potsdam, Germany\\
$^{2}$ Université de Lyon, Université Claude Bernard Lyon 1, CNRS/IN2P3, IP2I Lyon, F-69622, Villeurbanne, France \\
$^{3}$ Aix Marseille Université, CNRS/IN2P3, CPPM, Marseille, France \\
$^{4}$Institut de Recherche sur les Lois Fondamentales de l’Univers, CEA, Universit\'e Paris-Saclay, 91191 Gif-sur-Yvette, France \\
$^{5}$Institute for Astronomy, University of Hawaii, Honolulu HI 96822, USA\\
$^{6}$ Racah Institute of Physics, Hebrew University, Jerusalem 91904, Israel
\end{affiliations}

\parindent 0pt

\begin{abstract}
~ \\
Structure in the Universe is believed to have evolved out of  quantum fluctuations seeded by inflation in the early Universe. These fluctuations lead to density perturbations that grow via gravitational instability into large cosmological structures. In the linear regime, the growth of structure is directly coupled to the velocity field since perturbations are amplified by attracting (and accelerating) matter.
Surveys of galaxy redshifts and distances allow one to infer the underlying density and velocity fields. Here, assuming the \LCDM\ standard model of cosmology and applying a Hamiltonian Monte-Carlo algorithm to the grouped Cosmicflows-4 (CF4) compilation of 38,000 groups of galaxies, the large scale structure of the Universe is reconstructed out to a redshift corresponding to $\approx30,000\,\kms$. 
Our method provides a probabilistic assessment of the domains of gravitational potential minima: basins of attraction (BoA). 
Earlier Cosmicflows catalogs suggested the Milky Way Galaxy was associated with a BoA called Laniakea.
Now with the newer CF4 data, there is a slight probabilistic preference for Laniakea to be part of the much larger Shapley BoA.  The largest BoA recovered from the CF4 data is associated with the Sloan Great Wall with a volume within the sample of $15.5~10^6 (\Mpch)^3$, which is more than twice the size of the second largest Shapley BoA .
~\\
\end{abstract}

\begin{multicols}{2}
Coarse maps of the distribution of galaxies can be constructed by measuring redshifts.
However, the redshift of a galaxy, $z$, is a combination of the cosmic expansion velocity and its peculiar motion, $cz = H_0d + v_{pec}$, with $c$ the velocity of light, $d$ the distance, $H_{0}$ the Hubble constant (chosen to be consistent in the mean with distance measures), and $v_{pec}$ denotes a galaxy's deviation from pure Hubble flow in the radial (line of sight) direction. 
 Extensive surveys have been carried out over the last decades, cataloging the redshifts of millions of galaxies (\eg{} CfA\cite{DeLapparent1986}, SDSS\cite{York2000} and 6dF\cite{Jones2009} surveys among others). Such redshift surveys, using the approximation $d \simeq cz/H_0$, have uncovered an intricate network of clusters, filaments, sheets and voids - the so-called cosmic web. 
The regions of highest density delineated by rich clusters provided early three-dimensional maps of the skelaton of structure within $z=0.1$.\cite{Tully1992,Einasto1994}

\indent
However, redshift maps give an incomplete picture. First of all, not all matter is luminous: most of the gravitationally active matter in the Universe is ``dark matter'' with no electromagnetic manifestation\cite{Rubin1980}. In addition, galaxies are {\it biased} tracers of the density field\cite{Kaiser1987}, sampling the density field in a non-linear manner. The bias - namely the relation between galaxy number and the matter density - is non linear and varies with the morphological type of the galaxy. Moreover, counts of galaxies are sensitive to observational effects such as the Malmquist bias. 

\indent
Galaxy peculiar velocities, on the other hand, are unbiased tracers of the underlying matter density field because a galaxy's peculiar velocity (more aptly termed gravitational velocity) responds to the entire distribution of matter in the Universe, directly accounting for redshift distortions and independent of hidden mass biases. Hence we opt here to reconstruct the large scale structure from surveys of galaxy distances and peculiar velocities.  

\indent
The requirement of measuring a galaxy redshift is straightforward but the measurement of distance is challenging. An extensive catalog of distances (and hence peculiar velocities) like the catalog used here is inevitably a heterogeneous combination of various surveys with varying sensitivities and selection functions.
The Cosmicflows-4 catalog\cite{Tully2023} (CF4), used here, is the most extensive catalog to date comprising some 56,000 galaxies. These are gathered in a composite catalog of 38,000 constraints on the (linear) velocity field, composed of individual and groups of galaxies. Details of the components of the catalog are described in the Method supplement. Succinctly, there is good all-sky coverage outside the zone of Galactic obscuration to redshift $cz=15,000\kms$, with an extension of good coverage in the Galactic and celestial north to $cz=30,000\kms$. 

\indent
Since surveys of galaxy peculiar velocities provide only their radial (line of sight) component, we are left with the problem of reconstructing the full 3D field. Under the assumption of potential flow, the 3-dimensional velocity and density fields can be inferred from such data (e.g. the POTENT method\cite{Bertschinger1989}). Yet, galaxy peculiar velocities are noisy, sparse and incomplete and consequently the application of a POTENT-like method has a limited scope. A typical Cosmicflows data point has a relative error of about 20\% in its measured distance, which corresponds to a $\approx 2,000\kms$ uncertainty on the peculiar velocity for a galaxy at a redshift of $d \approx 100\Mpch$ ($cz\approx10,000\kms$). For the \LCDM model\cite{Planck2016} this corresponds to a signal to noise ratio of about 15\%;  a very noisy signal. A Bayesian approach is needed to overcome such a ``bad'' data - by incorporating the standard cosmological model as a prior. The Wiener filter (WF) and constrained realizations (CR) algorithm\cite{Bertschinger1989,Hoffman1992,Zaroubi1995} provides the optimal linear reconstruction tool from such data. Indeed, the WF/CRs method has been applied to reconstructions with peculiar velocity data\cite{Zaroubi1999,Doumler2013A,Sorce2015,Hoffman2021}.

\indent
Recently, forward, non-linear Monte Carlo modeling\cite{Kitaura2009,Jasche2013,Wang2014,Lavaux2016,Graziani2019}, in the context of Bayesian analysis, has proven to be a successful mathematical tool to reconstruct realizations of the underlying density and velocity fields that are consistent with the input data and the assumed Bayesian prior model. 

\indent
The HAmiltonian Monte carlo reconstruction of the Local EnvironmenT\cite{Valade2022} (HAMLET) method used in this work is such an approach. The method seeks to recover the Fourier modes of the density and velocity field, by assuming a cosmological model, here \LCDM\cite{Planck2016}. The exploration of the parameter space is performed with a Hamiltonian Monte Carlo\cite{Neal2011} (HMC) sampling algorithm and produces chains of states, where each of these states represents a possible reconstruction of the velocity and density field consistent with the constraining data, their errors and the assumed \LCDM\ model. The HAMLET method and its testing against mock data was presented elsewhere\cite{Valade2022,Valade2023}. 

\indent
In the following discussion we will focus on the particular feature of structures that we call a ``basin of attraction" (BoA). 
The entire Universe can be considered a patchwork of abutting BoAs, just as the terrestrial landscape is separated into watersheds. A BoA is generally not gravitationally bound, as the relative motion of distant points within it are usually dominated by the cosmic expansion. The volume filling factor of BoAs is unity, hence their mean density is the cosmic mean. Streamlines diverge out of the local maxima of the velocity potential and converge onto its local minima - namely they `stream' away from the underdense to the dense regions of the Universe. 
In co-moving coordinates, streamlines (defined in Method) represent the trajectories along which matter flows. Any source position in a BoA leads via a streamline to a ``sink'' near the potential minimum within the BoA. 
Thus a BoA, associated with a given local minimum of the potential, is the volume that encompasses all the streamlines that converge to a common region associated with a potential local minimum.

\indent
While the definition of a BoA is precise if distances and velocities are known without error, the data that informs their boundaries are noisy and incomplete.  
In previous studies\cite{Tully2014, Dupuy2023}, BoAs were computed on the mean inferred velocity field. This procedure is problematic as the construction of BoAs is not a linear process: the mean BoA from different realizations of the velocity field is not equal to the BoA built from the mean velocity field. An issue with previous BoA reconstructions based on the mean velocity field is that at the edge of the data BoAs may ``infinitely leak'' outwards, as the smoothed velocity field converges to the null field. This deficiency calls for a more elaborate method, presented here for the first time.

\indent
Each HMC trial gives links between sources and sinks consistent with the  data.  Because of imperfections in the data, the peculiar velocity fields and thus the basins of each state in the HMC chain differ from one realization to the next, while all being equally likely realizations of the velocity field constrained by the data and the \LCDM\ model. A BoA can thus be constructed for each state, resulting in an ensemble of BoAs from which the probabilistic Basin of Attraction (p-BoA) is constructed. A detailed description of the construction of the p-BoA's is given in the Method supplement. It is just briefly noted here that the core of the p-BoA is defined by the distribution of the sinks of the ensemble of BoAs. The probability that a volume element is associated with a given p-BoA is given by fraction of streamlines stemming from the volume element that terminate in this p-BoA. Our approach enables us to give two distinct probabilities; the probability that a basin of attraction exists and the probability for each sampled point of space to belong to it. Given the probabilistic nature of the constructed BoAs some are robustly predicted with sharply defined boundaries while others are less secure. Volume elements may be part of several p-BoAs, with probabilities summing to less than unity. As will be described, there is an uncertainty to the boundary of the BoA that we live in. 

\indent
Our results are presented by means of an accompanying video that provides a visualization of the density and velocity structure of the volume surveyed by CF4 following from our HMC analysis.  The figures in this article are key frames from the video.  The video begins with a display of the 3D distribution of galaxies/clusters sampled by CF4, transitioning to a vector representation of the measured radial peculiar velocities for the sample (blue inward, red outward).  Then we show density and velocity field reconstructions from individual HMC trials before showing the averaged density field over many realizations (see Fig.~1).  The particularly prominent Sloan Great Wall, Shapley, Hercules and South Pole Wall structures are identified.
More detailed features of high and low density are identified in a $4,000\kms$ thick slice on the supergalactic equator (SGZ=0), where known elements of the cosmography such as the 4-clusters or Sculptor Wall are indicated~\cite{Courtois2013, Pomarede2017}. 

\indent
The video then transitions to a display of the velocity streamlines from the mean reconstruction (see Fig.~2). There are a multitude of sinks of the mean streamlines.  The network of streamlines leading to each sink gives definition to an individual BoA. Near the mid-point of the accompanying video, one sees the separate BoA defined by streamlines encased in shells of different colors (see Fig.~3).  

\indent
The reader is invited to explore the 3D large scale structure in the accompanying interactive model, where individual data points - galaxies and groups of galaxies -  in the CF4 sample are shown as black points, an iso-surface in gray captures the reconstructed high density regions and five particularly interesting BoAs are encased (with distinctive colors) at surfaces given 50\% probability from the chain of 1000 realizations. These five highlighted BoAs include Shapley in yellow, a BoA centered in Ophiuchus associated with Laniakea in orange, Perseus-Pisces in cyan, Hercules in green, and Sloan Great Wall in red.

\indent
In the latter part of the video, ambiguities in BoA boundaries are explored from probabilities arising from the individual realizations.  The sink locations of scattered sources are shown, with the major aggregations of these identified with BoA (see Figs.~4 and 5). A given source can be associated with a given BoA with a probability garnered from the many trials.  

\indent
Streamlines sourced at the Milky Way find their way to two alternate sink regions.  The most favored terminus is in the region of the Shapley concentration\cite{Raychaudhury1989, Scaramella1989} (48\% ending in a well defined aggregate of sinks, enlarged to 58\% with inclusion of dispersed but proximate sinks). The remaining roughly 40\% of streamlines terminate in an aggregate in the vicinity of the Ophiuchus cluster.  The associated BoA in this case very much resembles the Laniakea BoA identified with data from CF2\cite{Tully2014}.

\indent
Within $cz=15,000\kms$ where there is good all-sky coverage (outside the zone of obscuration) large scale structures of high density found in earlier studies are recovered with generally improved acuity: the Shapley concentration at $cz\approx15,000\kms$\cite{Scaramella1989, Raychaudhury1989}, the Center for Astrophysics (CfA) Great Wall at $cz\approx6,000\kms$\cite{DeLapparent1986} and South Pole Wall at $cz\approx 11,000\kms$\cite{Pomarede2020}, the Perseus-Pisces filament at $cz\approx 6,000\kms$\cite{Haynes1988}, and the Hercules complex at $cz\approx9,000\kms$\cite{Einasto2001}. 
Nearby, evidence emerges for a BoA centered in proximity to the highly obscured Ophiuchus cluster that lies behind the center of the Milky Way Galaxy\cite{Wakamatsu1981, Hasegawa2000} at $cz \approx 9,000\kms$.  This BoA may include the so-called Great Attractor region\cite{Dressler1987b} and the entity Laniakea\cite{Tully2014} including ourselves.  
In the extension to $cz \approx 30,000\kms$, the Sloan Great Wall\cite{Gott2005} and associated structure is overwhelmingly dominant; indeed it is the most prominent structure so far mapped in detail in our reconstruction. 

\indent
Our methodology identifies 37, 15 and 7 p-BoAs with intrinsic probabilities of more than 20\%, 50\% and 75\%, respectively. We name each basin after the over-density which hosts the aggregation of sinks from the HAMLET trials. For example, a plurality of sinks that lie within the high density contours of the Shapley concentration gives definition to the Shapley p-BoA. 
In Extended Data, Table 1 records probabilities for the realities of the most consequential nearby BoA and Table 2 provides the location of the cores and of the associated volumes of the 15 p-BoAs with probabilties greater than 50\%, plus two of particular interest: Funnel and CfA Great Wall. The largest discovered p-BoA is the one associated with the Sloan Great Wall, encompassing a volume that is more than twice as great as the next biggest one, Shapley. 

\indent
Giving specific attention to the BoA called Laniakea from the analysis of CF2 data\cite{Tully2014},
two points are salient. The first is that this basin appears in only 62\% of the realizations, with an attractor in the highly obscured region of the Ophiuchus cluster. Meanwhile, the Milky Way is contained by this basin only about 40\% of the time. In fact, the Milky Way tends to more often be associated with the Shapley concentration (almost 50\% of the time tightly and almost 60\% of the time loosely). 
This leaves us in the  interesting situation of supporting the likely existence of an Ophiuchus/Laniakea p-BoA with only $\sim40\%$ probability of hosting our home galaxy. This option is explored in the illustration of Fig\,6. However, it is more probable that the Milky Way and indeed a large volume of Laniakea are within the domain of the Shapley p-BoA.

\indent
There is a similar uncertainty with the status of the CfA Great Wall\cite{DeLapparent1986}. The highest probability ($\sim40\%$) linkage is to the Shapley BoA; alternatively with probability $\sim 1/3$ is linkage to the Hercules BoA; and alternately with probability $\sim 1/4$ is a stand-alone CfA Great Wall BoA. There is this 3-way ambiguity even in a domain of dense data.

\indent
Given the finite depth of Cosmicflows data and the linear increase of uncertainties with distance, the robustness of the reconstruction of the velocity field diminishes at the edge of the data. The dominant BoAs identified here: Shapley, Hercules, and the Sloan Great Wall (and South Pole Wall) are all limited by the edges of our data. 
The current data are not deep enough to determine the outer bounds of these dominant p-BoAs. It follows that from the point of view of the BoAs that cosmology has not yet reached its ``end of greatness''.

\section*{Methods}
\label{sec:CF4:data}

\paragraph{Data} 
 
Cosmicflows-4 is a collection of 56,000 galaxy distances gathered into 38,000 groups\cite{Tully2023}. All told, there are eight distance measure methodologies, although the great bulk come from two methodologies involving correlations between global galactic photometric and kinematic properties: the luminosity-rotation rate relation for spiral galaxies\cite{Tully1977} (Tully-Fisher relation -- TFR) and the luminosity-velocity dispersion-surface brightness relation for elliptical galaxies\cite{Djorgovski1987, Dressler1987} (Fundamental Plane -- FP).  In turn, most of the contributions for these two methodologies come from three sources.  The largest contribution of 34,000 FP distances is confined to the sector north of the plane of the Milky Way and in the north celestial hemisphere and extends to $z=0.1$ in redshift\cite{Howlett2022}.  A second FP contribution of 9,000 distances is confined to the celestial south and redshift $z=0.05$\cite{Springob2014, Qin2018}.  The third major source, that of 10,000 TFR distances\cite{Kourkchi2022}, extends across the sky, mostly within $z\approx0.04$. The TFR material, dependent on kinematic information from radio telescopes,  is particularly plentiful at celestial latitudes  0$^{\rm o}$ to +38$^{\rm o}$ accessed by the large Arecibo Telescope, deficient below latitude $-45^{\rm o}$ accessed only by the smaller Parkes Telescope, and given intermediate coverage otherwise with the Green Bank Telescope. The resultant coverage of the sky with the ensemble of contributions is uneven.  Overall, there is reasonable coverage outside the zone of Milky Way obscuration across the sky within $z=0.05$ with a slight deficiency south of the Milky Way in the celestial north.  Then there is the substantial extension evident in the figures north of the Milky Way and in the celestial north. 

\indent
Scatter in distance and velocity measurements can be averaged over members of a galaxy group or cluster\cite{Tully2015, Tempel2017}, reducing errors. 
Each galaxy in CF4 is cross-matched as to whether it lies within the region of collapse spatially and in velocity of a cataloged group.  Groups can have from hundreds of distance measures (and independently, velocity measures) to just one.  Most galaxies in CF4 stand alone in unassigned groups.  The resulting catalog contains 38,000 entries.  Each entry consists of two angles defining positions on the sky and the group averaged redshifts and distance moduli with uncertainties.  Unsurprisingly, uncertainties are lowest nearby and in rich groups where the density of information is greatest.

\paragraph{Linear Theory} The complex statistics of the evolved universe can only be reproduced through elaborate
simulations. For this work we adopt the simpler modeling given by linear theory that holds where deviations from the homogeneous and isotropic Friedmann model are small. In this
context, the (over-)density field $\delta$ and the peculiar velocity field $\vv$ are bound by 
\begin{equation}
    - H_0 f(\Omega_m) \delta = \vnabdot \vv.
    \label{eq:cont}
\end{equation}
This equation can be solved in either direction, \ie{} the density can be obtained from the velocity field and
vice-versa. The fields follow the statistics of Multi Normal Variables, and thus random
realizations of the density field can be readily constructed from a matter power spectrum. In this
work, we consider the one provided by the Planck mission\cite{Planck2016}. In the \LCDM\ standard model linear theory is
generally considered an accurate description of the velocity field for scales above  $\approx 5 \Mpc/h$ (where $h=H_0/100$).
This sets the lower limit to the resolution of our reconstructions. 

\paragraph{Bayesian model} The \emph{posterior} probability is constructed as stated by the Bayes
theorem applied to the reconstruction of the (linear) density field $\delta$ and the recovery of the distance
of velocity tracers $\mathcal{D}$ from coupled measurements of redshifts $\mathcal{Z}$ and distance
moduli $\mathcal{M}$
\begin{equation}
    P(\delta, \mathcal{D} | \mathcal{Z}, \mathcal{M}) = 
    P(\mathcal{Z}, \mathcal{M} | \delta, \mathcal{D}) P(\delta, \mathcal{D}).
\end{equation}
The details of these functions are given a previously published method paper\cite{Valade2022}. Errors
on the distances moduli are assumed to be normally distributed and given in the CF4 catalog. Errors on the redshifts are also
assumed to be normally distributed with a fixed uncertainty of $50\kms$ augmented with the variance of the
marginalized non-linear part of the signal not recovered by the linear fields. We set the amplitude
of the non-linearities at a value of $\snl=100\kms$. This handles naturally the log-normal bias\cite{Hoffman2021,Sorce2023}, as well as the homogeneous Malmquiest bias. The inhomogenenous Malmquist bias and selection effects are partially corrected and are the focus of ongoing works. The model used in the work has been enhanced with respect to previous discussions\cite{Valade2022, Valade2023}: the selection function in redshift present in the data have been modeled\cite{Straus1995, Hinton2017}, and the computation of luminosity distances takes the peculiar movements of galaxies into account\cite{Calcino2017}.

\paragraph{Hamiltonian Monte Carlo} The parameter space is explored with a Hamiltonian Monte Carlo
method\cite{Neal2011,Betancourt2017}. The purpose of using a Monte Carlo chain is to produce a series of
independent realizations which obey a probability law. This is achieved by an iterative process, in which
each realization is constructed from the previous one. The variety of Monte Carlo methods
essentially derives from the variety of algorithms to perform this nuclear Monte Carlo step. The
Hamiltonian Monte Carlo method proposes a new realization by integrating the Hamilton equations of motion
in a potential defined by $-\log P$ for a particle whose initial position is the current realization
and whose initial momentum is picked at random. This method proves to be extremely efficient in
highly dimensional spaces, where other methods tend to stall. The meta parameters of the
exploration are tuned by the combined application of the No U-Turn Sampler (NUTS\cite{Hoffman2011})
and the Dual Averaging Step Adaptation\cite{Hoffman2011}. 

\paragraph{Streamlines and Basins of Attraction} Streamlines are a graphical visualization of a velocity field. The equation of 'motion' of the line element of a given stream line ${\bf s}({\bf l})$, where ${\bf l}$ is the line parameter, is:
\begin{equation}
    \frac{{\rm d}\,{\bf s}}{{\rm d}\,{\bf l}}={\bf v}({\bf l})
    \label{eq:line}
\end{equation}
The choice of the seeding points of the streamlines is a matter of convention. Particles move along streamlines at a given moment, yet they only traverse a small fraction of a given streamline in the age of the Universe.

\indent
All the cells whose stream line converges to the same point in space, called the
attractor, define a Basin of Attraction (BoA).
The study of BoAs in a cosmological simulation\cite{Dupuy2020} shows that these
structures are stable in time, although a small basin may be absorbed by larger neighbors.The adopted smoothing scale similarly might lead to the merger of neighboring basins. 

\indent
Comment on terminology: in earlier work\cite{Tully2014, Pomarede2015} we gave the term `supercluster' the same meaning we now give for BoA. Note that the term ``watershed supercluster'' has also been used\cite{Dupuy2023} and the related concept of "supercluster cocoons"\cite{Einasto2020}. However, confusion arises because there is no community consensus on the meaning of the term supercluster.  Lacking an agreed definition, we avoid calling any structure a supercluster.

\paragraph{Probabilistic basins of attractions} The reconstruction of the flow field is done within a Bayesian framework, in which the Hamiltonian Monte Carlo Markov Chain constructs states that sample the posterior probability distribution function, given the data and the assumed prior \LCDM\ model. The HAMLET algorithm employed here produces an ensemble of such states, from which an ensemble of BoAs is constructed. The statistical distribution of the ensemble is consistent with the posterior probability. We extend here the definition of a given BoA of a given state to a statistical one whose probability reflects the distribution of the HMC states.  
We consider the ensemble of all sinks of streamlines of all the states, excluding the sinks that lie outside of the volume covered by the data. 
We identify regions densely populated by sinks with a machine learning clustering algorithm, HDBSCAN\cite{Malzer2019}. This algorithm identifying aggregates of sinks and outperforms a simpler friends-of-friends method by its ability to select a density threshold per aggregate and the stability of its results with respect to its hyperparameters. Aggregates that are too small or too low-density are disregarded, while the remaining aggregates define the core of the so-called probabilistic basins of attractions (p-BoA). We define the intrinsic probability of a p-BoA by the proportion of realizations in which it appears. The probability of a given source location on the grid to belong to a given p-BoA is defined by the proportion of realizations in which this source converges to a sink position within a given aggregate.

\newpage
\section*{Data Availability}

\href{https://vimeo.com/963076278/4d5ed63fd6}{\bf Thirteen minute video} following the discussion in the text and illustrating the properties of large scale structure in the volume of the Universe sampled by Cosmicflows-4.

\href{https://vimeo.com/963076278/4d5ed63fd6}{https://vimeo.com/963076278/4d5ed63fd6}

\href{http://sketchfab.com/3d-models/pboas-p05-cf4-mean-field-delta-and-velocity-9ba49209e60c48de8469b01ee8ee772e}{\bf Interactive model} illustrating the distribution of the Cosmicflows-4 sample of galaxies, an iso-contour of the mean Hamiltonian Monte Carlo density reconstruction, mean velocity streamlines, and the 50\% probability inclusion shells of five prominent basins of attraction.

The estimated density and velocity fields are available upon reasonable request to the authors. We also offer an online tool that correlates distance and velocities using the constructed velocity field within this program. Please access the tool at \href{http://edd.ifa.hawaii.edu/CF4calculator/}{http://edd.ifa.hawaii.edu/CF4calculator/}, and consult with the instruction on how to use the tool and interpret the results \cite{Kourkchi2020-DVcalculator}. We will continuously update the calculator for better visualizations and to incorporate the latest improvements.

\section*{Data and code availability}

The observational data utilized in this study was placed in the public domain in connection with the publication of Cosmicflows-4\cite{Tully2023}.

The code used to produce this work is not public but can be communicated in response to reasonable requests. 



\section*{Author Contributions Statement}

AV developed the key ideas of this work, designed the reconstruction of the density and velocity field, as well as the computation of the streamlines and finally participated in the writing of the paper. NIL, YH and SP took active parts in the development of the methods described in the paper, as well as its redaction. AV, DP and RBT analyzed the cosmography. DP generated all the visualizations (figures and videos) accompanying this work. RBT led the writing of the paper and was deeply involved in designing its science goals. EK built the online CF4 distance calculator.

\end{multicols}

\label{table:volume}

\clearpage
\section*{Figures}

\begin{figure*}[h!]
 \includegraphics[width=\textwidth]{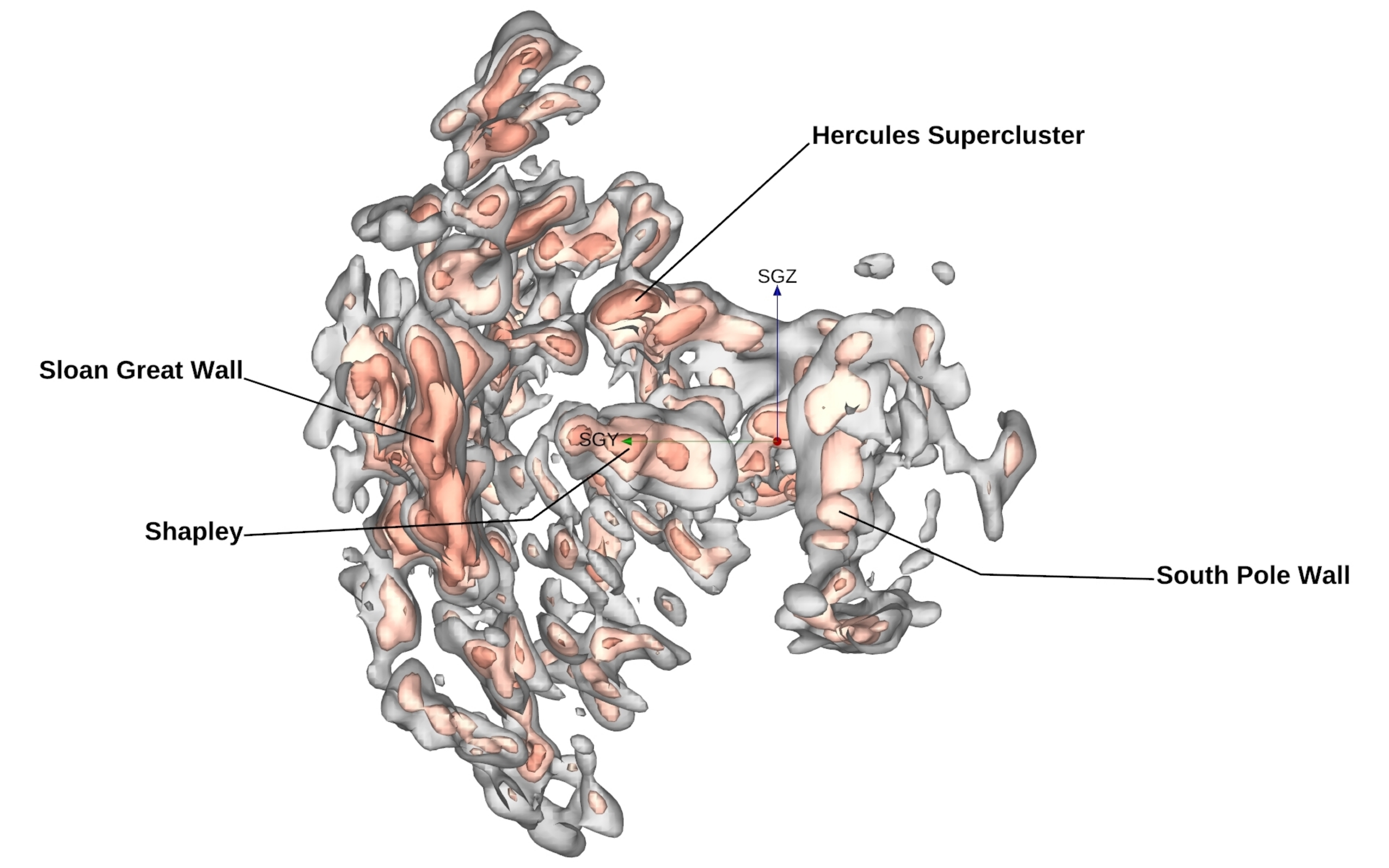}
  \caption{High density perturbations $\delta$ in the distribution of matter in the Local Universe with the mean of many HMC trials. The gray and red contours enclose regions of increasing density. Prominent structures appearing in this view are labeled. The coordinate system is supergalactic and the Milky Way Galaxy is at the origin of the 10,000 km/s long red (SGX), green (SGY), and blue (SGZ) axes.  This view is looking in along the red axis from negative SGX. Video frame at 02:08.}
	\label{fig:1}
\end{figure*}

\clearpage

\begin{figure*}[h!]
	\begin{center}
  \includegraphics[width=\textwidth]{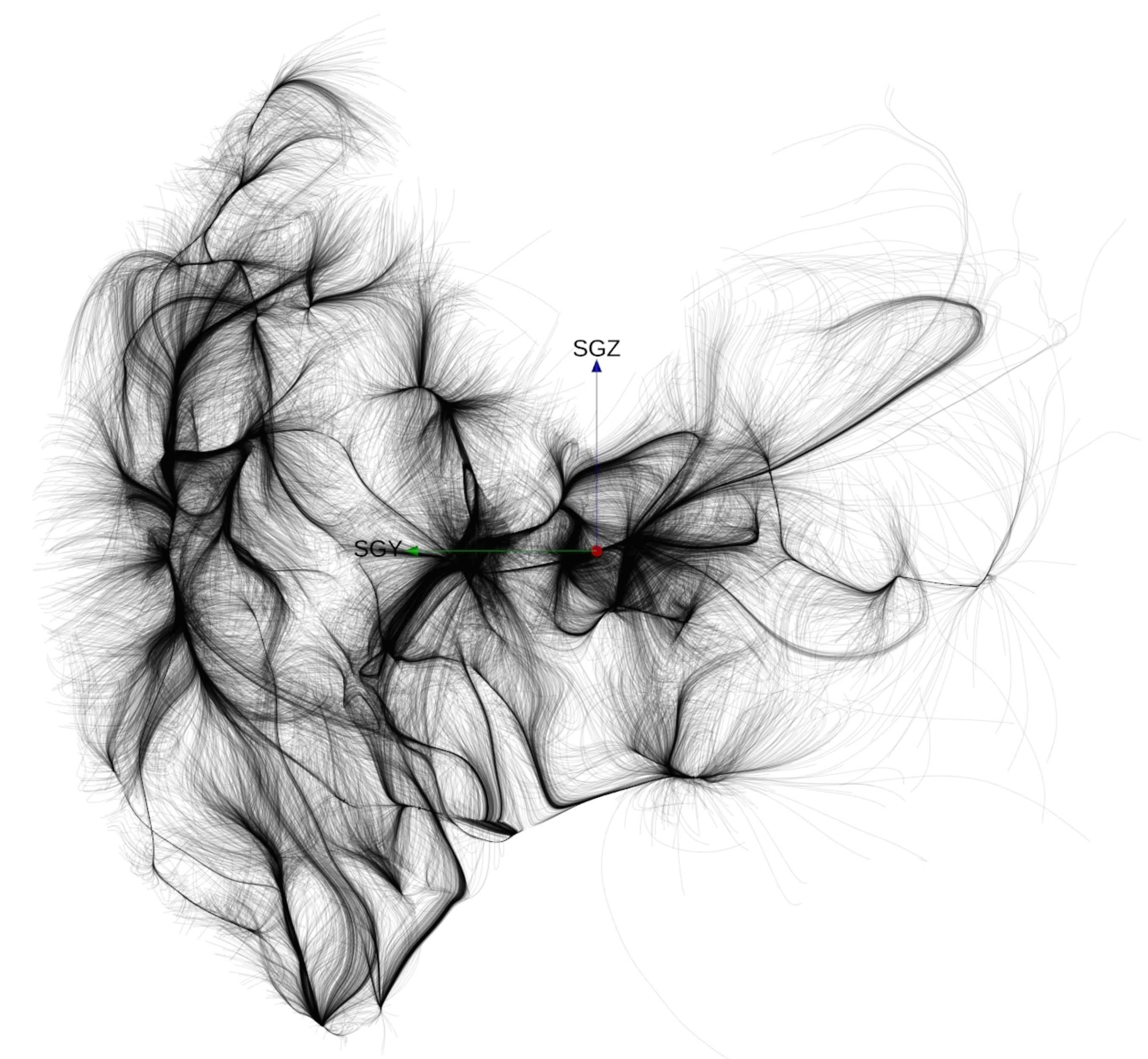}
  \caption{Streamlines constructed from the mean HMC velocity field.  Looking in from negative SGX with the Milky Way at the origin of the red, green, blue axes. Video frame at 03:34.}
	\end{center}
	\label{fig:2}
\end{figure*}

\clearpage

\begin{figure*}[h!]
  \includegraphics[width=\textwidth]{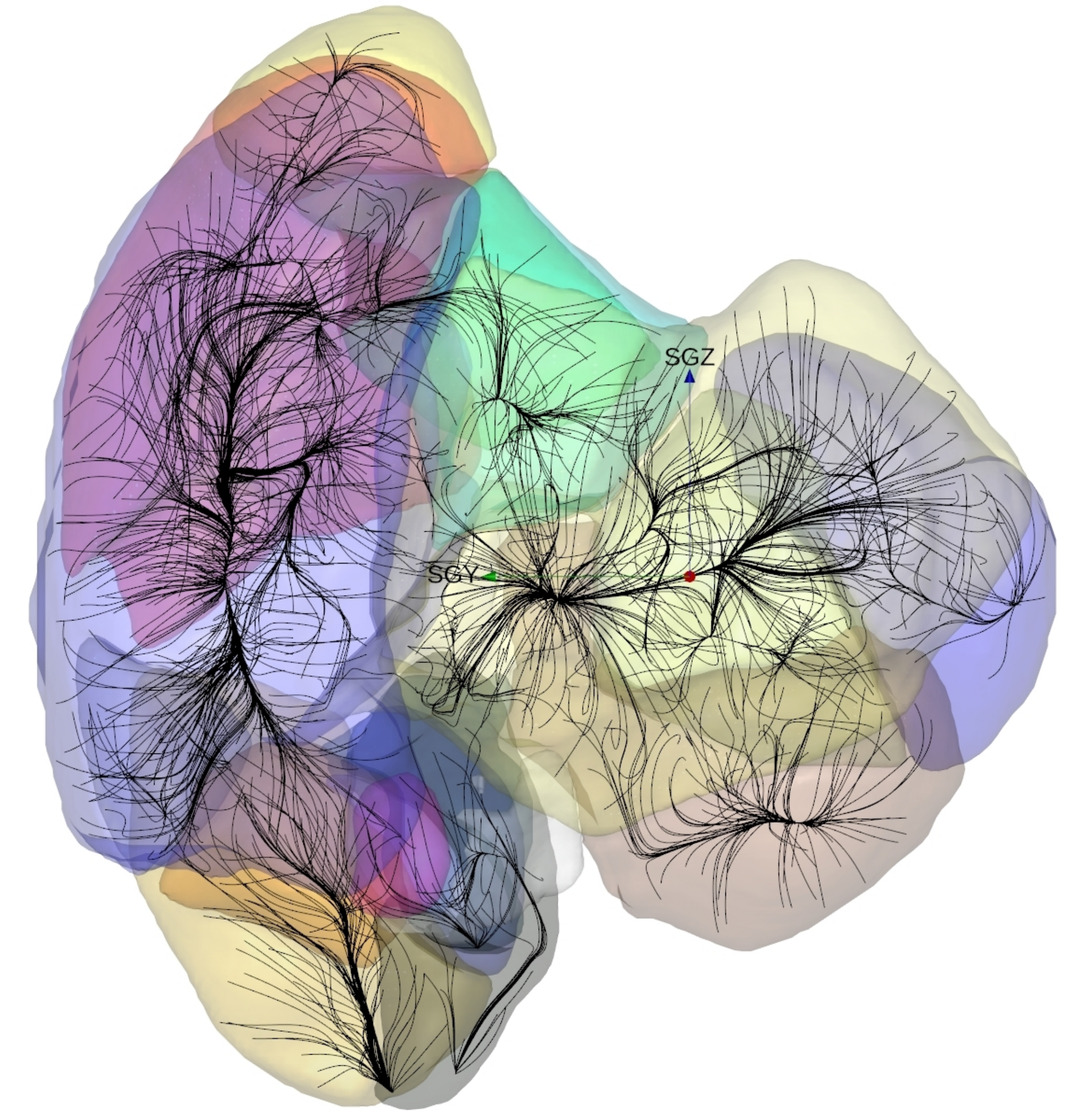}
  \caption{Velocity streamlines seeded at arbitrary locations within the reconstructed volume, with colored envelopes associated with separate BoA extracted from the mean field. To counter the "infinite leaking effect", these BoAs have been cropped to the region covered by data. Looking in from negative SGX. Video frame at 04:18.}
	\label{fig:3}
\end{figure*}

\clearpage

\begin{figure*}[h!]
  \includegraphics[width=\textwidth]{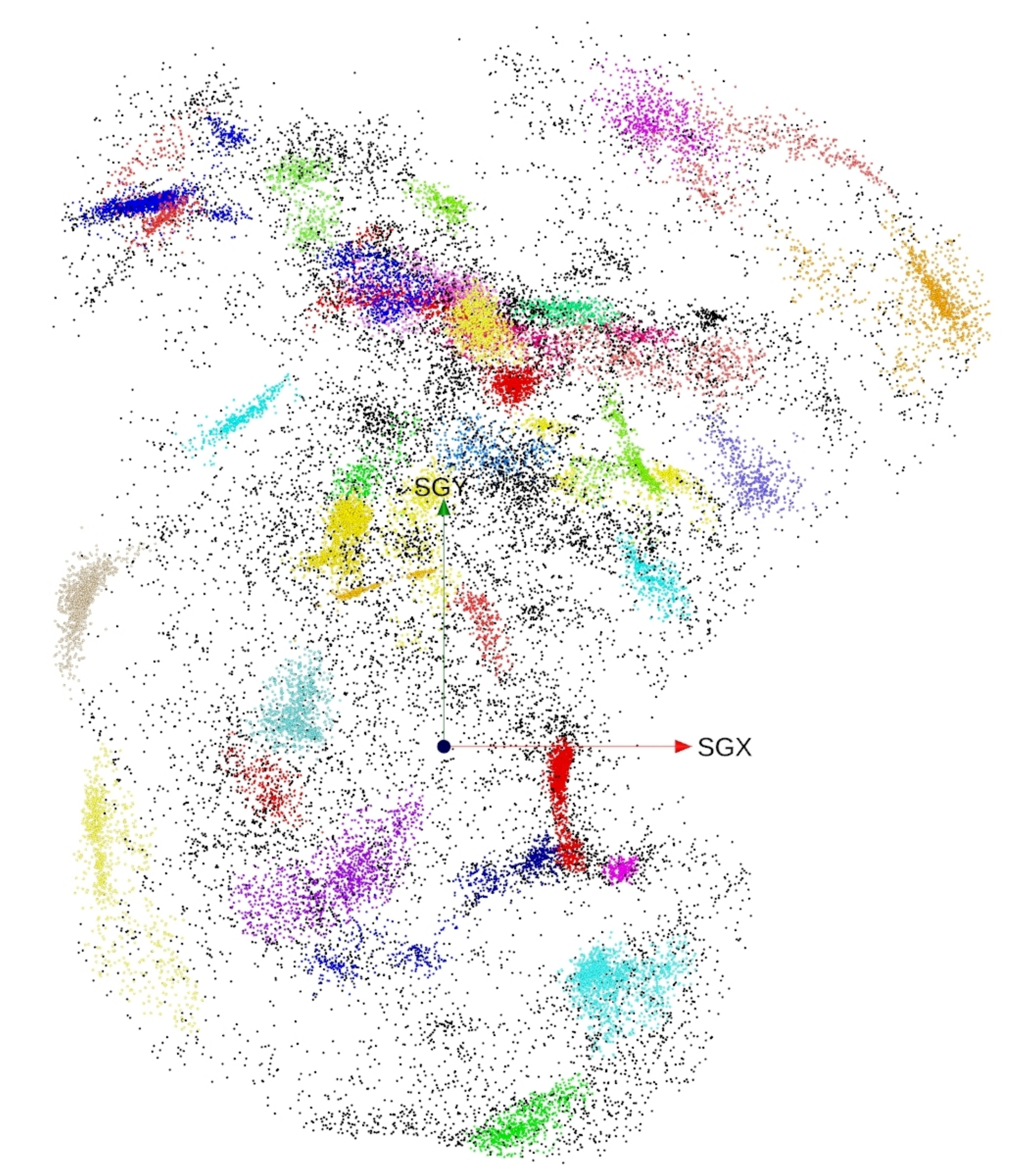}
  \caption{Sinks of velocity streamlines from individual HMC trials.  Colors are associated with the major aggregates of streamline end points. Looking in from positive SGZ. Video frame 06:58.}
	\label{fig:4}
\end{figure*}

\clearpage

\begin{figure*}[h!]
  \includegraphics[width=\textwidth]{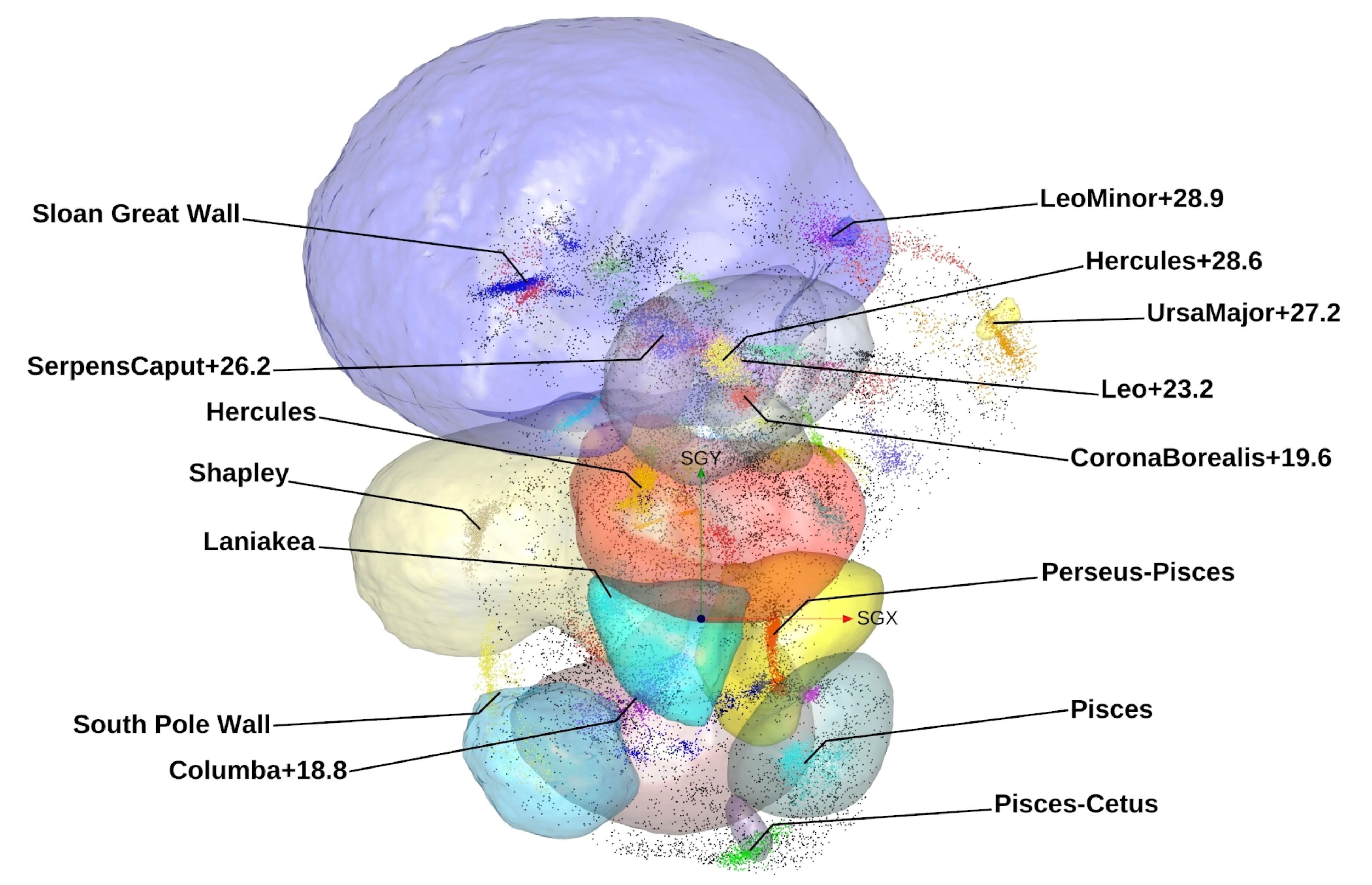}
  \caption{Envelopes of major BoA superimposed on the sinks of HMC trials shown in the previous figure.  Names given to the major BoA. For previously unnamed objects, we use the convention "constellation+distance" in units of 1,000 km/s. Note that several structures can be found in the same constellation (\eg{} Hercules). Video frame at 07:44.}
	\label{fig:6}
\end{figure*}

\clearpage

\begin{figure*}[h!]
  \includegraphics[width=\textwidth]{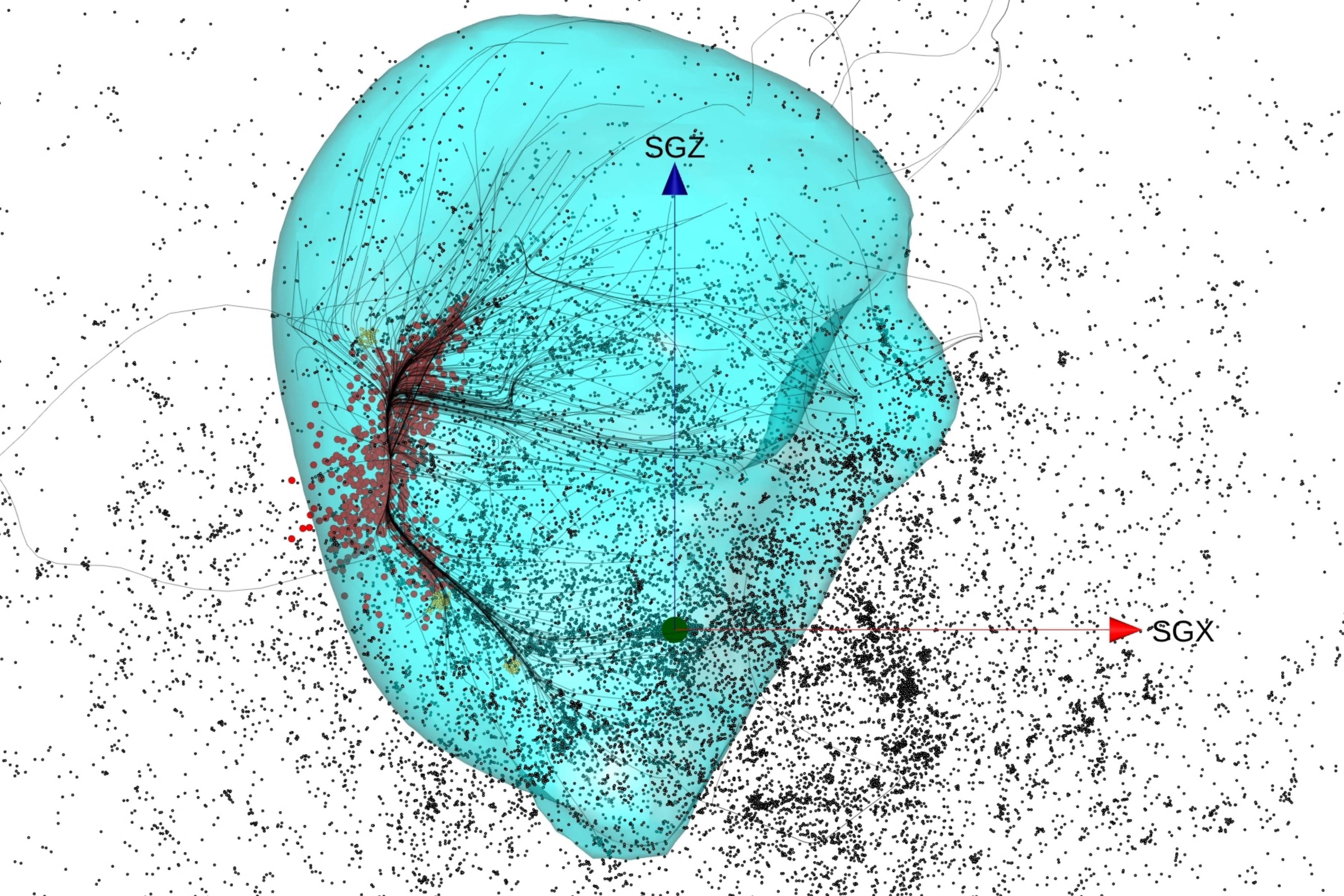}
  \caption{
  Cyan surface envelopes Ophiuchus/Laniakea BoA at 20\% probability.  Sinks from multiple trials shown as red points. Streamlines are from one of the realizations.  Galaxies from the CF4 group sample are shown if $-9500<\rm{SGY}<+4500$~\kms. Axes have lengths 10,000~\kms.
  Video frame at 09:37.
  }
	\label{fig:7}
\end{figure*}

\clearpage

\begin{table*}[h!]
\begin{tabular}{lcccccccccccccccccccccccccccccccccccc}
                      & \rot{Ophiuchus BoA (Laniakea)} & \rot{Shapley BoA} & \rot{South Pole Wall BoA} & \rot{Hercules BoA} & \rot{CfA Great Wall BoA} & \rot{Perseus BoA} & \rot{Funnel BoA} &\rot{Other} \\
\hline
Milky Way             &          \cmap{39} (38) & \cmap{58} (48)             &                    \cmap{1} & \cmap{0}    & \cmap{0}                &                    \cmap{0} & \cmap{0} & \cmap{2} (12)      \\
Shapley core (A3558)  &                \cmap{0} & \cmap{99} (89)             &                    \cmap{0} & \cmap{0}    & \cmap{0}                &                    \cmap{0} & \cmap{0} & \cmap{1} (11)       \\
Coma                  &                \cmap{0} & \cmap{40} (35)             &                    \cmap{0} & \cmap{33}   & \cmap{25} (23)          &                    \cmap{0} & \cmap{0} & \cmap{2} (11)      \\
South Pole Wall       &                \cmap{0} & \cmap{21}                  &                   \cmap{42} & \cmap{0}    & \cmap{0}                &                    \cmap{0} & \cmap{0} & \cmap{35}          \\
Hercules supercluster &                \cmap{0} & \cmap{0}                   &                    \cmap{0} & \cmap{99}   & \cmap{0}                &                    \cmap{0} & \cmap{0} & \cmap{1}           \\
Perseus cluster       &                \cmap{0} & \cmap{0}                   &                    \cmap{0} & \cmap{0}    & \cmap{0}                &                    \cmap{92} & \cmap{0} & \cmap{8}           \\
Funnel (A400)         &                \cmap{0} & \cmap{0}                   &                    \cmap{0} & \cmap{0}    & \cmap{0}                &                    \cmap{72} & \cmap{6} & \cmap{22}          \\
\end{tabular}
\caption{Probability (in \%) of some known features of the Local Universe to be part of probabilistic basins of attraction. Four claimed superclusters of the Sloan Great Wall\cite{Einasto2001} converge to the entity SCl\,126 with a probability of about 95\%. The percentages in brackets or alone given in this table are obtained by the automatic aggregation of attractor sinks by a machine learning algorithm. The larger values are  based on a visual inspection in the 3D model of the locations of the scattered sinks with respect to the large aggregations of points.} 
\end{table*}

\begin{table*}[h!]
    \small
        \begin{tabular}{l|c|c|c|c|c|c|c|c}
            \hline
            \multicolumn{2}{c}{}                 & \multicolumn{6}{|c|}{Core position }                     &                \\
                             & Prob & SGX              & SGY              & SGZ               & R.A.  & Dec & cz & Volume \\
            Name             & \%   & \!$\Mpch$        & \!$\Mpch$        & \!$\Mpch$         & deg   & deg & km/s & $10^6 (\Mpch)^{3}$ \\
            \hline
            
            Sloan Great Wall  & 99   &-114.6 $\pm$ 15.0 & 221.0 $\pm$  9.1 &   8.5 $\pm$ 22.0  & 195.9 &  -0.4 & 24909 & 15.51 \\
            Hercules          & 99   & -39.6 $\pm$  5.7 &  86.4 $\pm$  8.6 &  79.5 $\pm$  6.5  & 232.3 &  11.4 & 12391 &  1.86 \\
            Perseus cluster   & 93   &  47.8 $\pm$  2.7 & -18.3 $\pm$ 14.6 & -32.7 $\pm$  4.1  &  61.1 &  23.8 &  6074 &  1.06 \\
            Shapley           & 90   &-145.1 $\pm$  5.8 &  59.1 $\pm$ 10.7 & -12.2 $\pm$  8.7  & 201.6 & -41.0 & 15715 &  7.02 \\
            Columba+13.8      & 89   & -43.1 $\pm$ 18.6 & -52.6 $\pm$ 12.8 &-119.8 $\pm$  3.5  &  88.0 & -42.0 & 13774 &  2.69 \\
            Pisces            & 87   &  68.1 $\pm$ 11.5 & -94.4 $\pm$  8.4 &  20.3 $\pm$ 25.3  &   6.6 &  11.6 & 11817 &  0.66 \\
            Leo+23.2          & 76   &  27.2 $\pm$ 27.4 & 169.2 $\pm$  8.4 &-156.5 $\pm$ 11.7  & 141.9 &  13.4 & 23207 &  0.51 \\
            Hercules+28.6     & 72   &  14.7 $\pm$  6.3 & 169.8 $\pm$  7.3 & 229.3 $\pm$ 10.0  & 247.3 &  31.1 & 28569 &  4.11 \\
            South Pole Wall   & 66   &-132.1 $\pm$ 11.5 & -48.6 $\pm$ 26.7 &  18.7 $\pm$ 23.0  & 268.4 & -65.9 & 14200 &  3.40 \\
            UrsaMajor+27.2    & 64   & 192.1 $\pm$ 18.3 & 180.7 $\pm$ 13.0 & -67.0 $\pm$ 45.8  & 133.6 &  55.7 & 27210 &  0.48 \\
            Ophiuchus         & 62   & -59.4 $\pm$  7.0 &  14.6 $\pm$  9.3 &  38.6 $\pm$ 16.1  & 245.3 & -28.5 &  7233 &  0.80 \\
            CorBor+19.6       & 61   &  28.4 $\pm$  5.1 & 145.7 $\pm$  3.8 & 127.4 $\pm$  7.1  & 233.6 &  38.4 & 19564 &  1.67 \\
            SerpensCaput+26.2 & 58   & -25.1 $\pm$ 10.6 & 185.9 $\pm$  8.4 & 183.5 $\pm$ 10.0  & 235.8 &  24.2 & 26237 &  2.19 \\
            Pisces Cetus      & 57   &  32.4 $\pm$ 11.0 &-151.4 $\pm$  7.3 & -20.5 $\pm$ 12.6  &  17.3 & -15.8 & 15621 &  0.81 \\
            LeoMinor+28.9     & 55   &  86.5 $\pm$ 12.1 & 250.9 $\pm$  8.9 &-113.5 $\pm$ 14.5  & 153.0 &  31.6 & 28868 &  0.54 \\
            Funnel            & 24   &  71.7 $\pm$  3.2 & -49.6 $\pm$  2.6 & -37.5 $\pm$  3.7  &  45.6 &  18.7 &  9491 &  1.02 \\
            CfA Great Wall    & 23   & -28.4 $\pm$ 12.7 &  64.1 $\pm$  3.9 &  11.9 $\pm$ 10.1  & 202.4 &   5.0 &  7111 &  0.61 \\

        \end{tabular}
        \normalsize
        \caption{Properties of the p-BoAs appearing in Table 1. Each core position is the mean of the ensemble of sinks from realizations that constitutes the p-BoA. The associated  volume is the mean of the volumes of each realization of the p-BoA and the error bars represent the scatter over the ensemble of states. The p-BoAs are sorted by their statistical significance.}
\end{table*}

\clearpage

\section*{References}


\end{document}